\documentclass[sigconf]{acmart}
\renewcommand\footnotetextcopyrightpermission[1]{}
\usepackage{amsmath}
\usepackage{graphicx}
\usepackage{subcaption}
\usepackage{booktabs}
\usepackage{tabularx}
\usepackage{xspace}
\usepackage{etoolbox}
\AtBeginEnvironment{table}{\footnotesize\setlength{\tabcolsep}{3pt}\renewcommand{\arraystretch}{0.92}}
\AtEndEnvironment{table}{\normalsize}
\AtBeginEnvironment{table*}{\footnotesize\setlength{\tabcolsep}{3pt}\renewcommand{\arraystretch}{0.92}}
\AtEndEnvironment{table*}{\normalsize}

\makeatletter
\renewcommand{\@fnsymbol}[1]{%
  \ensuremath{%
    \ifcase#1\or \dagger\or \mathsection\or *\or \ddagger\or  \mathparagraph\or \|\or **\or \dagger\dagger\or \ddagger\ddagger \else\@ctrerr\fi%
  }%
}
\makeatother

\newcommand{\proj}{\textit{SolarChain}\xspace}

\AtBeginDocument{%
  }

\copyrightyear{2026}
\acmYear{2026}
\setcopyright{cc}
\setcctype{by}
\acmConference[UbiComp Companion '26]{Companion of the 2026 ACM International Joint Conference on Pervasive and Ubiquitous Computing}{October 11--15, 2026}{Shanghai, China}
\acmBooktitle{Companion of the 2026 ACM International Joint Conference on Pervasive and Ubiquitous Computing (UbiComp Companion '26), October 11--15, 2026, Shanghai, China}
\acmDOI{10.1145/3798063.3839423}
\acmISBN{979-8-4007-2533-3/2026/10}

\begin{document}

\title{SolarChain: A Physics-Grounded Embodied IoT System for Verifiable Urban Solar Market Design}

\newcommand{\sharedaffiliation}{%
  \affiliation{%
    \institution{Duke Kunshan University}
    \city{Suzhou}
    \country{China}
  }%
}

\newcommand{\jhuaffiliation}{%
  \affiliation{%
    \institution{Johns Hopkins University}
    \city{Baltimore}
    \state{Maryland}
    \country{United States}
  }%
}

\author{Shilin Ou}
\sharedaffiliation
\authornote{Equal contribution.}

\author{Yifan Xu}
\sharedaffiliation
\authornotemark[1]

\author{Zhenshan Zhang}
\sharedaffiliation
\authornotemark[1]
\authornote{Z.Z. completed this work while at DKU and is now at Johns Hopkins University.}
\author{Luyao Zhang}
\sharedaffiliation
\authornote{Joint corresponding authors. \\ Email: Zhang:  lz183@duke.edu; Huang: mh596@duke.edu. Address: 8 Duke Ave, Kunshan, Suzhou, Jiangsu 215316, China.}

\author{Ming-Chun Huang}
\sharedaffiliation
\authornotemark[3]

\renewcommand{\shortauthors}{Ou et al.}

\begin{abstract}
Distributed solar markets must coordinate physical reports, economic allocation, and public settlement even when IoT data can be manipulated. We present \proj, a controlled Embodied Intelligence of Things (EIoT) prototype that integrates four functions: physics-bounded screening of photovoltaic reports, persistent agent and planner coordination, configurable allocation between producer rewards and market liquidity, and replayable hash-linked auditing of settlement decisions. The benchmark combines city-level historical weather inputs with physics-modeled generation bounds and synthetic nodes, demand, trades, and scripted attacks. On 36,000 monthly records, an IQR/MAD baseline attains F1=1.000, while the rule-based adaptive verifier attains F1=0.988 and provides physically interpretable decision evidence; it is not uniformly superior across attack classes. A sensitivity sweep selects a 20/80 reward/liquidity default under the stated simulation assumptions, while showing the incentive--liquidity trade-off. These results provide reproducible evidence from a controlled prototype, not proof of deployment readiness. We release the code, data, and audit artifacts in the spirit of open science.

\end{abstract}



\begin{teaserfigure}
    \centering
    \includegraphics[width=\textwidth]{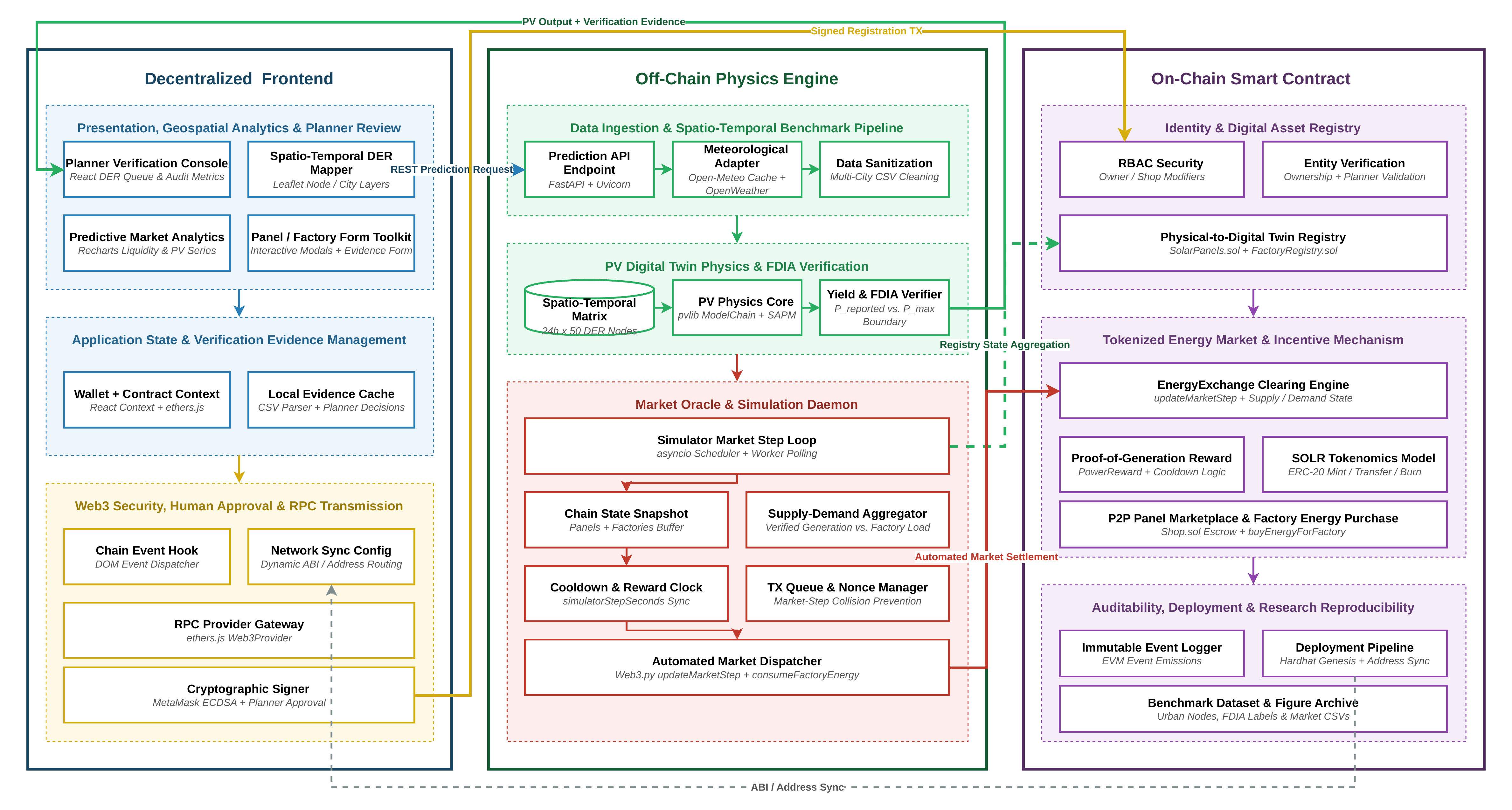}
    \caption{System architecture of \proj. The three-layer implementation connects an off-chain physics engine, a decentralized frontend, and on-chain contracts. Market allocation is configurable, with a 20/80 producer-reward/liquidity default in this prototype.}
    \Description{System architecture diagram connecting the decentralized frontend, off-chain physics engine, and on-chain smart contracts through verification evidence, registration transactions, market settlement, and address synchronization flows.}
    \label{fig:system_architecture}
\end{teaserfigure}

\maketitle

\section{Introduction}

Urban photovoltaic (PV) systems increasingly participate in local energy coordination rather than operating as passive generators, adding flexibility and management complexity to distributed grids \cite{liang2025managing,aoun2024centralized}. Peer-to-peer markets can reward prosumers and aggregate distributed flexibility \cite{abdella2018peer,morstyn2018peer}, but they also couple physical measurements to digital incentives. A ledger can preserve a submitted record without establishing that the underlying generation report was physically plausible. False-data injection can therefore create fictitious supply, rewards, or settlement events \cite{liu2011false,musleh2023spatio}. Market design introduces a second dependency: assigning more accepted value to individual rewards leaves less value for shared liquidity.

Prior studies address grid anomaly detection, on-chain energy markets, and decentralized physical infrastructure as distinct problems. Their combination leaves a systems question: \emph{Can a controlled Embodied Intelligence of Things (EIoT) prototype connect physics-bounded PV reports to coordinated market settlement with traceable evidence, and what limitations become visible when that chain is evaluated end to end?} This question concerns integration and evidence, not the novelty of each component in isolation.

We answer this question with \proj. City-level historical weather and solar geometry first bound each simulated PV report. Persistent Solar, Planner, Factory, and Market agents then coordinate verification and participation. A configurable contract divides accepted value between producer rewards and market liquidity, while hash-linked events make each decision replayable. In EIoT terms \cite{liu2026eiot}, the simulated PV body, weather context, and physical constraints constitute \emph{enactment}. Cross-role coordination provides \emph{engagement}, and accumulated residuals with updates to calibration, trust, and participation state represent \emph{evolution}. This evolution is rule-based rather than learned continual adaptation. Because no physical agents are deployed, the study evaluates a simulated decision path rather than a field system.

This paper makes three system-level contributions:
\begin{itemize}
    \item a cross-layer workflow linking bounded PV reports, persistent coordination, configurable allocation, and contract settlement;
    \item a replayable evaluation spanning statistical and physics-grounded verifiers, 11 scripted attack classes, allocation sensitivity, runtime, and audit-chain integrity; and
    \item an open benchmark separating sourced weather, physics-modeled variables, and synthetic system behavior so the evidence can be reproduced without overstating field validity.
\end{itemize}

The results are deliberately scoped to a simulated prototype. The attacks are scripted rather than adaptive, and the inputs contain neither inverter nor smart-meter telemetry. Code, data, and audit artifacts are available online.\footnote{\url{https://github.com/Global-Nomad-Nexus/SolarSave}}

\section{Background and Design Requirements}

\subsection{EIoT and Decentralized Infrastructure}
EIoT characterizes embodied systems through enacted grounding, engaged interaction, and evolutionary adaptation \cite{liu2026eiot}. Urban computing supplies the broader spatial and infrastructural setting \cite{zheng2014urban}, while DePIN research studies how distributed contributors can provide physical resources under digital coordination and incentives \cite{ballandies2023taxonomy,lin2024depin}. These perspectives motivate a stateful, cross-layer system, but they do not by themselves authenticate a PV measurement. We therefore treat EIoT as an organizing framework and explicitly distinguish simulated embodiment from deployment.

\subsection{Physical Plausibility and False Data}
Conventional residual tests can miss attacks constructed to preserve a grid estimator's expected residual \cite{liu2011false}; spatiotemporal and learned detectors add contextual evidence \cite{musleh2023spatio,mohammadi2021detecting,parvez2025graph}. Physics-informed methods encode domain constraints into inference~\cite{raissi2019physics}, while secure-control frameworks protect distributed operation against false-data injection~\cite{dai2023blockchain}. For PV systems, established models connect irradiance, solar position, temperature, and hardware parameters to expected output \cite{skoplaki2009temperature,holmgren2018pvlib,anderson2023pvlib}. Such an envelope can reject impossible reports without training data, but a plausible within-bound value may still be false. A useful verifier should thus retain multiple forms of evidence and expose rather than conceal its attack-class failures.

\begin{figure*}[t!]
  \centering
  \begin{subfigure}[t]{0.49\textwidth}
    \centering
    \includegraphics[width=\linewidth]{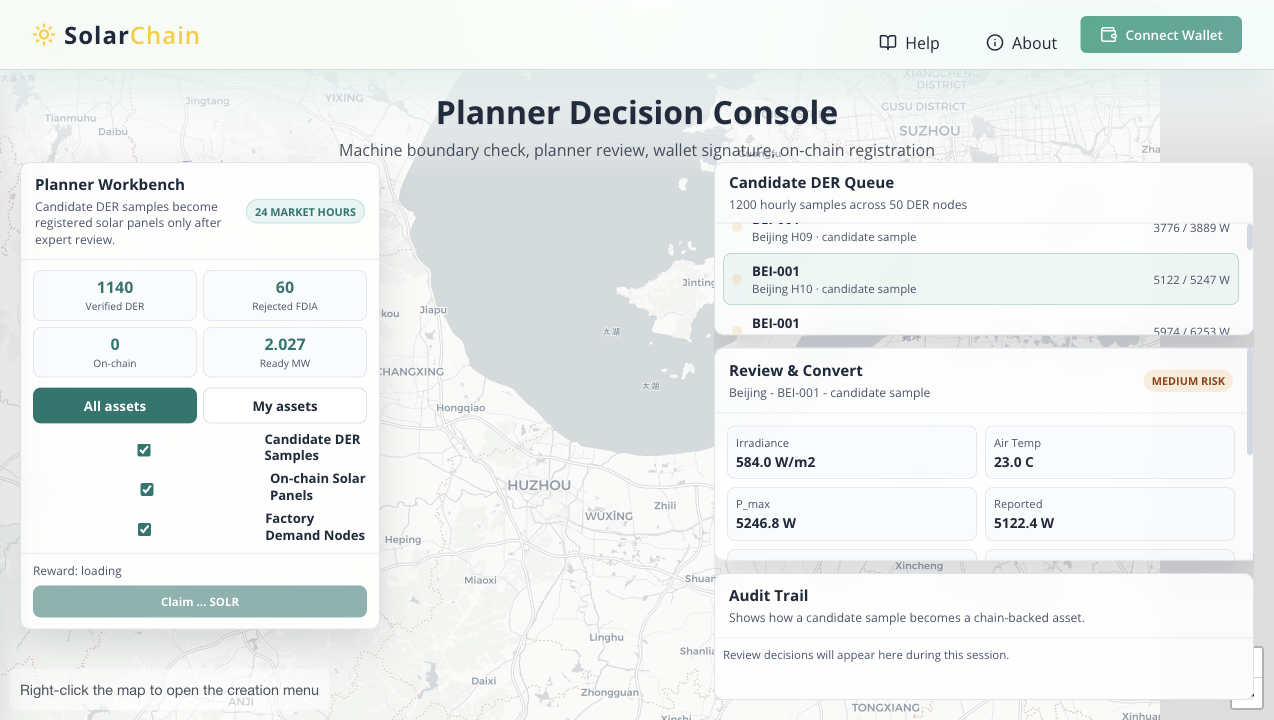}
    \caption{Planner console for reviewing candidate reports and bounds.}
    \label{fig:demo-planner}
  \end{subfigure}
  \hfill
  \begin{subfigure}[t]{0.49\textwidth}
    \centering
    \includegraphics[width=\linewidth]{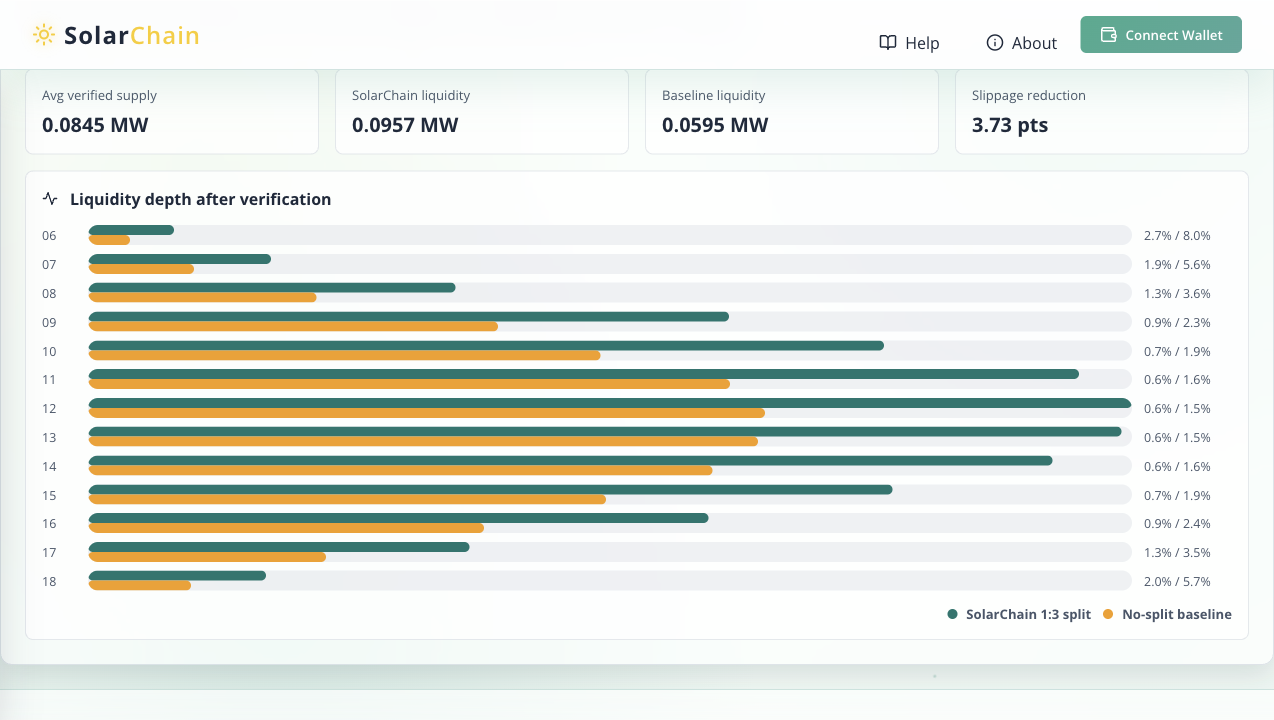}
    \caption{Market dashboard for comparing allocation policies.}
    \label{fig:demo-market}
  \end{subfigure}
  \caption{Demonstration interfaces in the \proj prototype.}
  \Description{Two prototype screenshots: a planner console for reviewing distributed energy reports and a market dashboard for comparing liquidity policies.}
  \label{fig:solarchain-demo}
\end{figure*}

\begin{table}[t!]
\centering
\caption{Mapping Between EIoT dimensions and the controlled prototype.}
\label{tab:eiot-mapping}
\begin{tabularx}{\columnwidth}{@{}lX@{}}
\toprule
\textbf{Dimension} & \textbf{Prototype summary (realization; evidence; boundary)} \\ \midrule
Enacted & PV nodes + city weather + physics-bounded reports; 50 nodes and 36{,}000 node-hours with modeled $P_{\max}$; no inverter/meter telemetry. \\
Engaged & Solar/Planner/Factory/Market roles coordinate via reports, demand, and settlement events; decision + market traces, local-EVM, hash-linked log; rule-based trust. \\
Evolutionary & Residual, trust, calibration, and participation state persists across steps; state histories and attack-wise verifier results; stateful but not learned adaptation. \\
\bottomrule
\end{tabularx}
\end{table}

\subsection{Energy Markets, Ledgers, and Oracles}
Peer-to-peer energy markets coordinate prosumer exchange and flexibility \cite{abdella2018peer,morstyn2018peer}; blockchain implementations make market state and rules inspectable \cite{mollah2020blockchain,huang2024p2p,sahih2024blockchain}. However, blockchain energy-trading systems still depend on off-chain measurements, and validation and oracle mechanisms manage rather than eliminate that trust boundary~\cite{westphall2022blockchain,xian2024distributed}. Liquidity mechanisms also rest on explicit price and allocation assumptions \cite{angeris2020improved}. Renewable ledgers must additionally address participant identity and double counting \cite{ferdous2023ssi}. Accordingly, a hash-linked settlement trace proves what the software processed, not the truth of the originating sensor or the environmental meaning of a token event.

Together, these strands motivate a decision chain in which physical evidence informs coordination, accepted supply enters a configurable market, and settlement remains replayable. \proj exposes how failures propagate across this chain.

\section{Integrated System Design}

\subsection{Physical Evidence and Stateful Coordination}
The off-chain engine combines a node's area $A$, efficiency $\eta$, effective irradiance $G$, temperature coefficient $\beta$, and reference temperature $T_{ref}$ to compute
\begin{equation}
P_{\max}=A\eta G[1-\beta(T-T_{ref})].
\end{equation}
A report above $\tau P_{\max}$ is physically inconsistent. Reports inside this envelope may still be flagged by temporal, neighbor, or residual-memory rules. A persistent SolarAgent retains residual history, calibration, trust, and participation state; the PlannerAgent approves, rejects, or escalates; and Factory and Market agents return demand and slippage feedback. The resulting coordination is inspectable, but it is neither a learned policy nor cryptographic proof of source authenticity.

State evolves only through declared transitions. The SolarAgent keeps the most recent 48 residuals, updates calibration after at least six observations, and revises trust after planner and market feedback. Threshold rules map trust to active, probation, or suspended participation. The PlannerAgent combines verifier outcome and trust to approve, reject, or escalate a report rather than silently overwriting it. This makes ``evolutionary'' testable as accumulated state without claiming online learning.

All calculations use one-hour intervals: $E_t=P_t\Delta t$ with $\Delta t=1\,\mathrm{h}$. Consequently, summed kW (MW) samples are reported as kWh (MWh). Legacy implementation fields ending in \texttt{W} or \texttt{MW}, including contract inputs, represent power-derived one-hour step values rather than separately calibrated energy-meter readings.

\subsection{Allocation, Settlement, and Replay}
Accepted supply enters the market contract, where reward share $\rho$ is configurable:
\begin{equation}
R_{reward}=\rho E_{verified},\qquad R_{liq}=(1-\rho)E_{verified}.
\end{equation}
The owner-set default is 2,000 basis points (20/80 reward/liquidity), an experimental policy parameter evaluated rather than assumed optimal. Wallet-signed contracts register assets, update supply and demand, burn SOLR for simulated factory purchases, and emit settlement events.

The event bus serializes each report, verification result, planner action, market update, feedback update, and settlement with the previous event hash. Replaying the chain reveals whether the recorded path has changed. The frontend in Figure~\ref{fig:solarchain-demo} presents this evidence before registration or purchase; physical calculations remain off-chain, and contracts record approved shared-state transitions.

This separation also exposes the trust boundary. Weather and node metadata feed the model, the planner approves its evidence, the contract enforces access and allocation bounds, and the ledger preserves accepted transitions. A later layer cannot repair a compromised input, although the combined record allows an auditor to identify the evidence used.

\section{Controlled Prototype Evaluation}

The evaluation examines three linked properties: detector accuracy and failure modes, sensitivity to the reward--liquidity allocation, and consistency from accepted reports through settlement and replay.

\subsection{Benchmark, Provenance, and Protocol}
The canonical benchmark covers April 1--May 1, 2026: 50 simulated PV nodes in five cities, 720 hourly observations per node, and 36,000 records in total. A fixed seed injects 1,800 labeled records into the original monthly FDIA set. The broader taxonomy contains this original category and ten additional scripted attack scenarios (11 categories in total); it does not model an adversary that observes the detector and adapts to evade it.

We distinguish three provenance classes. Externally sourced inputs are city-level historical temperature and shortwave-radiation series retrieved from Open-Meteo, not panel telemetry. Solar position, the clear-sky envelope, temperature correction, and $P_{\max}$ are physics-modeled. PV nodes and hardware parameters, reported power, demand, trades, planner/agent behavior, and attacks are synthetic. Dataset hashes, the weather cache, seed, scripts, and software versions accompany the release.

All detectors receive the same ordered records and labels. Precision, recall, and F1 use attacks as the positive class; FAR denotes the fraction of attacks accepted and FRR the fraction of normal reports rejected. Runtime uses 30 warmed full-batch runs on the environment named in Table~\ref{tab:verifier-main}. Because the benchmark is fixed-seed and deterministic, we report descriptive results rather than inferential significance. The taxonomy is a stress test, not a substitute for naturally occurring fault or adversarial data.

\subsection{Verifier Accuracy, Attack Classes, and Cost}
Table~\ref{tab:verifier-main} reports the monthly comparison. IQR/MAD achieves the highest aggregate F1 (1.000); the adaptive verifier reaches 0.988 and is therefore not uniformly superior. Its physics, temporal, neighbor, and residual/trust rules provide inspectable evidence and replayable state at the cost of higher runtime and occasional false positives.

Category results expose the boundary: nighttime, above-bound, and neighbor-inconsistent attacks are strongly covered; sensor drift, replay, and intermittent bursts are partially covered; and near-bound fraud, weather spoofing, and coordinated near-bound attacks remain failures. The released \texttt{attack\_taxonomy\_results.csv} reports seven detectors across all 11 categories. The 24-hour result (F1=1.000 on 60 obvious violations) is only a boundary-violation sanity check.

IQR/MAD is preferable when aggregate classification on the injected distribution is the sole objective. The adaptive verifier instead supplies a physical residual, temporal context, and recorded state transition at a median 52.159~ms per batch (1.449~$\mu$s per record). These explanations do not compensate for uncovered attack classes, which require authenticated context or additional detection mechanisms.

Separately, all 144,721 generated events reproduced their stored hash linkage. This establishes trace integrity for that run, not source truth or better detection.

\subsection{Allocation Sensitivity}
The reward share $\rho$ is configurable, with liquidity receiving $1-\rho$. The sweep evaluates 41 ratios in 2.5-percentage-point increments across seven common scenarios. Each ratio passes through the same demand-response and market-update path before averaging and scoring. This controls the simulated scenarios but cannot identify causal participant response because supply, demand, and behavior are model-generated. Table~\ref{tab:ratio-main} reports representative endpoints, the selected neighborhood, and the 50/50 reference; the released CSV contains every ratio. The default $\rho=0.20$ is conditional on these assumptions rather than a universal optimum.

\begin{figure*}[t!]
  \centering
  \includegraphics[width=\textwidth]{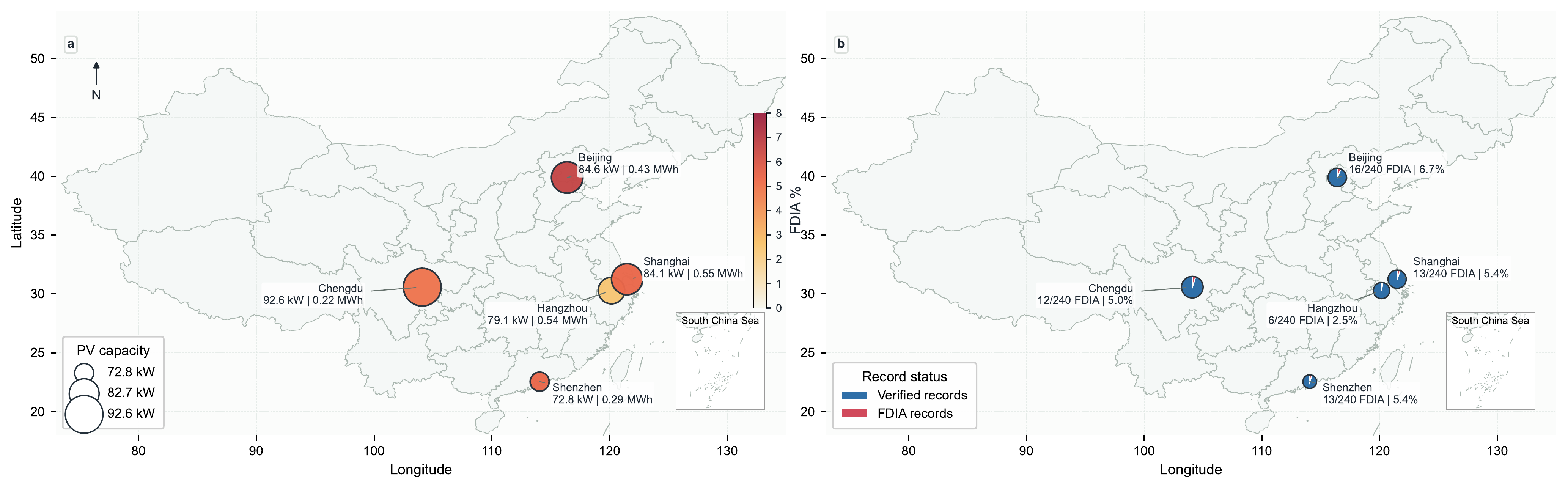}
  \caption{Spatial benchmark illustration for one 24-hour slice across five Chinese cities. Markers locate the 50 simulated PV nodes; the right panel summarizes the slice's 1,200 records and 60 scripted anomalies. The canonical evaluation extends the same pipeline to 36,000 April records.}
  \Description{Map of Beijing, Shanghai, Chengdu, Shenzhen, and Hangzhou with simulated photovoltaic nodes, accompanied by 24-hour generation and false-data injection summaries.}
  \label{fig:china-urban-fdia}
\end{figure*}

\begin{table*}[t!]
\centering
\caption{Verifier results on 36,000 monthly records with 1,800 original FDIA labels. FAR is false-acceptance rate and FRR is false-rejection rate. Runtime columns report the median and p95 of 30 warmed batch runs on the released macOS/Apple-arm64 environment.}
\label{tab:verifier-main}
\scriptsize
\setlength{\tabcolsep}{4pt}
\begin{tabular}{lrrrrrrrr}
\toprule
Detector & Precision & Recall & F1 & FAR & FRR & Median (ms) & p95 (ms) & $\mu$s/record \\
\midrule
3-sigma & 1.000 & 0.766 & 0.868 & 0.234 & 0.000 & 6.378 & 7.605 & 0.177 \\
IQR/MAD & 1.000 & 1.000 & \textbf{1.000} & 0.000 & 0.000 & 7.859 & 9.195 & 0.218 \\
Physics bound & 1.000 & 0.862 & 0.926 & 0.138 & 0.000 & 0.132 & 0.205 & 0.004 \\
Adaptive SolarAgent & 0.976 & 1.000 & 0.988 & 0.000 & 0.001 & 52.159 & 53.769 & 1.449 \\
\bottomrule
\end{tabular}
\end{table*}

\begin{table}[H]
\centering
\caption{Allocation sensitivity averaged across the simulated market scenarios. ``Reward/liquidity'' gives producer and market shares; score is unitless.}
\label{tab:ratio-main}
\scriptsize
\setlength{\tabcolsep}{2.2pt}
\begin{tabular}{lrrrrr}
\toprule
Allocation & \shortstack{Fulfill\\(ratio)} & \shortstack{Reward\\(kWh)} & \shortstack{Slip.\\(\%)} & \shortstack{Unmet\\(MWh)} & Score \\
\midrule
0/100   & 0.961 & 0     & 32.356 & 1.869 & 0.617 \\
10/90   & 0.964 & 4,174 & 32.165 & 1.686 & 0.640 \\
\textbf{20/80} & \textbf{0.956} & \textbf{8,976} & \textbf{32.569} & \textbf{2.111} & \textbf{0.651} \\
25/75   & 0.939 & 11,220 & 33.032 & 2.858 & 0.646 \\
50/50   & 0.782 & 22,441 & 33.297 & 9.520 & 0.592 \\
\bottomrule
\end{tabular}
\end{table}

The score uses declared weights (0.50 fulfillment, 0.20 reward, 0.15 slippage, 0.10 unmet demand, and 0.05 drawdown). Different weights or dynamics can select another ratio; the contract therefore exposes a bounds-checked basis-point parameter.

\subsection{End-to-End Spatial and Settlement Trace}
The May 1, 2026 interface trace contains 1,200 node-hour records and 60 deliberately obvious anomalies. Unlike the monthly detector benchmark, it follows accepted reports through city aggregates, allocation, purchases, and the event log.

\begin{table}[H]
\centering
\caption{Simulated city-level generation in the 24-hour companion trace.}
\label{tab:city-trace}
\scriptsize
\setlength{\tabcolsep}{3.2pt}
\begin{tabular}{lrrrr}
\toprule
City & Capacity (kW) & Verified (kWh) & Peak (kW) & Cap. factor \\
\midrule
Beijing  & 84.58 & 427.72 & 63.34 & 21.07\% \\
Chengdu  & 92.62 & 220.95 & 51.23 & 9.94\% \\
Hangzhou & 79.14 & 539.39 & 72.12 & 28.40\% \\
Shanghai & 84.12 & 554.34 & 76.32 & 27.46\% \\
Shenzhen & 72.77 & 285.73 & 46.26 & 16.36\% \\
\bottomrule
\end{tabular}
\end{table}

The five synthetic city clusters produce 2,028.13~kWh of accepted generation. Chengdu has the largest simulated capacity but the lowest verified output, whereas Shanghai and Hangzhou have smaller capacities and higher output. This difference follows city-level weather and synthetic hardware; it is not a field comparison of municipal PV fleets.

The market applies the 20/80 split after aggregating accepted supply, so rejected records cannot directly create reward or liquidity in this trace. Factory purchases consume simulated supply and emit the settlements in Table~\ref{tab:p2p-trace}; legacy 50/50 fields serve only as a reference policy.

\begin{table}[H]
\centering
\caption{Simulated P2P settlement outcomes in the 24-hour companion trace.}
\label{tab:p2p-trace}
\scriptsize
\setlength{\tabcolsep}{3.4pt}
\begin{tabular}{lrrrr}
\toprule
City & Trades & Volume (MWh) & SOLR burned & Exergy (MJ) \\
\midrule
Beijing  & 7  & 0.0994 & 96.74  & 10.84 \\
Chengdu  & 8  & 0.0883 & 89.13  & 7.45 \\
Hangzhou & 6  & 0.0855 & 85.21  & 7.05 \\
Shanghai & 14 & 0.2161 & 219.75 & 23.26 \\
Shenzhen & 7  & 0.1159 & 118.93 & 11.02 \\
\bottomrule
\end{tabular}
\end{table}

Across 42 synthetic purchases, the trace records 0.6052~MWh, 609.77 SOLR-equivalent units burned, and 59.63~MJ in the model's exergy field. Volume sums per-purchase MW values over one-hour steps. These values demonstrate software linkage, not real transactions, emissions, or carbon abatement.

\paragraph{Market-state comparison.}
Under common accepted generation, demand, reserve, and slippage assumptions, Table~\ref{tab:market-trace} compares the 20/80 policy with a simulated 50/50 reference (not a no-split market). More liquidity increases modeled depth and decreases the slippage proxy while assigning 0.4056~MWh-equivalent value to rewards.

\begin{table}[H]
\centering
\caption{Market-state metrics for the 24-hour companion trace.}
\label{tab:market-trace}
\scriptsize
\setlength{\tabcolsep}{3.3pt}
\begin{tabular}{lrrr}
\toprule
Metric & 20/80 default & 50/50 ref. & Change \\
\midrule
Average liquidity (MW) & 0.0856 & 0.0503 & +70.3\% \\
Peak liquidity (MW) & 0.2376 & 0.1453 & +63.6\% \\
Liquidity area (MW-hours) & 2.0545 & 1.2061 & +70.3\% \\
Average slippage proxy & 34.33\% & 57.18\% & -40.0\% \\
Daylight slippage proxy & 49.79\% & 82.15\% & -39.4\% \\
\bottomrule
\end{tabular}
\end{table}

These simulator outputs are neither exchange observations nor welfare estimates; the slippage proxy responds mechanically to synthetic demand and liquidity. Plausible attacks that evade verification could still alter market state, so field validation must jointly measure detection error and economic impact.

The trace checks that rejected generation is excluded, reward and liquidity sum to the accepted one-hour value, and every event names its preceding hash. Regenerated city and purchase totals satisfy these invariants, while local-EVM tests cover registration, access control, ratio bounds, token burning, and settlement fields. Full replay reconstructs the recorded inputs and decisions but remains a consistency test, not proof of contract security or source authenticity.

\section{Discussion and Limitations}

\subsection{Integrated Evidence}
The controlled evidence shows that an EIoT-oriented prototype can connect physical screening to coordinated settlement and replayable evidence. Enacted grounding rejects clearly impossible PV reports; engaged roles propagate approval and demand into market state; and evolutionary state retains residual, trust, calibration, and participation history. The result is an inspectable decision chain, not evidence that physics bounds, agents, contracts, or hash chains are individually novel.

The evidence also exposes a clear boundary. Physical plausibility is not authenticity: near-bound manipulation, coordinated reports, or spoofed weather can remain inside the rules. The adaptive verifier is therefore a reason-producing planner checkpoint, not a uniformly better detector; IQR/MAD has stronger aggregate F1 in the monthly benchmark. These results favor a layered comparison rather than a single winner. IQR/MAD can screen batch-level deviations, whereas the adaptive agent preserves residuals and state changes that explain a planner decision. Their outputs should remain distinct because agreement among rules derived from the same modeled envelope is not independent corroboration. Likewise, the 20/80 default is the highest-scoring point under this simulator's demand, participation, normalization, and declared weights, not a general welfare optimum.

A detector can classify well without supplying physical evidence for a settlement dispute; a market setting can improve a simulated score while relying on unverified supply; and perfect replay can preserve a false input. Linking the stages makes these dependencies visible and traces downstream effects to their assumptions. This distinction also changes what should be measured. Detection metrics quantify classification, whereas planner escalations, decision reversals, and excluded settlements quantify operational consequences. Reporting both avoids treating an explanatory trace as an accuracy gain or a high F1 score as sufficient market protection, and it allows each error to be followed into the allocation and settlement records it affects.

\subsection{Market, Audit, and Governance Implications}
A larger reward share strengthens direct producer incentives but removes value from shared liquidity; a larger liquidity share can support fulfillment while weakening individual reward. Exposing the split in basis points permits a deployment study to substitute observed behavior and an explicit governance process. Selecting $\rho$ should therefore be treated as a governance decision with published score weights, scenario ranges, and update intervals. A pilot should report the full trade-off among producer reward, fulfilled demand, unmet demand, slippage, and drawdown rather than only the highest composite score. Formal market and participant-response analysis would still be required before interpreting any selected ratio as policy.

The hash-linked log supplies a different kind of evidence from detector accuracy: it connects report, rule outputs, planner action, market update, and settlement so an auditor can reproduce a state change. It cannot prove sensor calibration, asset control, or environmental additionality. This distinction matters because digital incentives can become detached from physical activity \cite{alshater2026depin}, while renewable accounting still faces identity and double-counting risks \cite{ferdous2023ssi}. The off-chain process and planner remain trusted, and the contract owner can change the reward ratio; the prototype records these actions but implements neither decentralized oracle consensus nor stakeholder voting and dispute resolution.

Operational governance would therefore need accountable roles for meter attestation, model maintenance, approval and appeal, and allocation-policy control. Concentrating them in one operator preserves replayability but not institutional independence; decentralizing them without assigning responsibility may make failures harder to resolve. At minimum, model and ratio changes should have separate permissions and versioned justifications, while an appeal record should retain the parameters active when a report was judged. The prototype exposes these decisions but does not select a governance model.

\subsection{Threats to Validity and Validation Agenda}
Construct validity is limited by modeled rather than metered generation and by a slippage proxy rather than observed prices. Internal validity is constrained by synthetic hardware, demand, trades, agent behavior, and fixed-seed attacks. External validity is limited by one month of city-level weather and the absence of communication loss, clock drift, shading, clipping, outages, strategic bidding, privacy costs, chain congestion, and adaptive adversaries. Hash replay establishes event integrity but cannot authenticate an input at origin.

The perfect IQR/MAD result is distribution-specific: scripted anomalies can be unusually separable, and synthetic labels contain no field-data uncertainty. The adaptive verifier also depends on observation order and early feedback, so attack timing and cold-start behavior may change its result. Field labels should distinguish manipulation from shading, sensor drift, outages, and data loss; otherwise a single F1 score conflates adversarial behavior with operational faults. Category-level reporting reduces but does not remove these threats.

The release mitigates reproducibility concerns by recording the April window, seed, cached weather, file hashes, detector settings, scripts, and software versions. Runtime results also include the input hash and machine environment. Repeating the benchmark therefore does not require a live weather API, although it reproduces only the stated controlled conditions. An independent replication should reconstruct weather-derived bounds from the cache, reproduce event hashes, and compare regenerated metric tables with the released CSVs. Any divergence could then be localized to source data, software environment, or policy parameters.

Validation should next replay synchronized inverter and smart-meter traces through the same interface, then progress to shadow operation and a governed microgrid pilot. It should test seasonal faults, missing data, adaptive attacks, end-to-end latency, privacy and communication overhead, and the economic consequences of false acceptance and rejection. Chronological evaluation should separate cold-start from steady-state periods and propagate each false decision through settlement to quantify downstream value at risk. A field protocol must also define meter calibration, ratio governance, appeal procedures, and behavior during unavailable weather or network data before any claim of deployment readiness.

\section{Conclusion}
This study asked whether a controlled EIoT prototype could carry physics-bounded PV reports through coordinated settlement while retaining traceable evidence. \proj implements this path, but the results set clear limits. IQR/MAD achieved the best aggregate F1, while the interpretable adaptive verifier retained attack-class failures. The 20/80 allocation remained assumption-dependent, and hash replay established trace consistency rather than source authenticity.

\proj therefore serves as an experimental testbed for tracing how verification errors and allocation choices affect digital incentives. Field claims require synchronized telemetry, adaptive-attack evaluation, and governed trials.

\begin{acks}
Ming-Chun Huang and Luyao Zhang received research support from the Duke Kunshan University Education Development Foundation and the Kunshan Municipal Government. Shilin Ou was additionally supported by the Summer Research Scholar Program at Duke Kunshan University under the mentorship of Luyao Zhang.
\end{acks}

\bibliographystyle{ACM-Reference-Format}
\bibliography{references}
\appendix
\section*{Appendix}

\section{Benchmark and Verifier Specification}
\label{app:benchmark-verifier}
This appendix specifies the controlled benchmark and its evaluation artifacts.

\begin{table}[H]
  \centering
  \caption{Canonical benchmark specification.}
  \label{tab:appendix-benchmark}
  \normalsize
  \setlength{\tabcolsep}{3pt}
  \begin{tabularx}{\columnwidth}{@{}lX@{}}
    \toprule
    Item & Specification \\
    \midrule
    Window & 2026-04-01 to 2026-05-01; 720 hourly steps (Asia/Shanghai). \\
    Scale & Five cities; 50 simulated nodes; 36,000 node-hours. \\
    Labels & 1,800 original FDIA rows (5\%). \\
    Seeds & Dataset \texttt{20260511}; taxonomy \texttt{20260617}. \\
    Sourced & Open-Meteo city temperature and shortwave radiation. \\
    Modeled & Solar position, clear-sky envelope, temperature correction, $P_{\max}$. \\
    Synthetic & Nodes, reports, labels, demand, trades, agents, and attacks. \\
    SHA-256 & \texttt{caa0d95656e618e7d54c98bfed507128a}\newline
      \texttt{94edac454b8b964ac06f3541f9b190e}. \\
    \bottomrule
  \end{tabularx}
\end{table}

\begin{table}[H]
  \centering
  \caption{Released benchmark artifacts.}
  \label{tab:appendix-artifacts}
  \normalsize
  \setlength{\tabcolsep}{3pt}
  \begin{tabularx}{\columnwidth}{@{}l r X@{}}
    \toprule
    Artifact & Rows & Role \\
    \midrule
    Nodes & 50 & Registry and PV parameters. \\
    Generation & 36,000 & Bounds, reports, and labels. \\
    Market & 720 & Demand, allocation, and slippage. \\
    Trades & 1,185 & Simulated energy trades. \\
    Event log & 144,721 & Ordered agent and settlement events. \\
    \bottomrule
  \end{tabularx}
\end{table}
\noindent Files: \path{urban_energy_nodes.csv}.\par
\noindent \path{market_liquidity.csv}; \path{p2p_trades.csv}.\par
\noindent \path{spatiotemporal_generation.csv}.\par
\noindent \path{eiot_event_log.csv.gz}.

For the offline verifier, let $d_{i,t}=\max(P^{\max}_{i,t},1)$,
$r_{i,t}=(P^{\mathrm{rep}}_{i,t}-0.978P^{\max}_{i,t})/d_{i,t}$, and
$q_{i,t}=P^{\mathrm{rep}}_{i,t}/d_{i,t}$. Its physical flag is
\begin{equation}
 B_{i,t}=[P^{\mathrm{rep}}_{i,t}>1.03P^{\max}_{i,t}+25]
 \lor[P^{\max}_{i,t}\leq1\land P^{\mathrm{rep}}_{i,t}>50].
 \label{eq:appendix-physics-flag}
\end{equation}
The temporal rule $T$ compares $r_{i,t}$ with up to 12 prior node residuals
(minimum four; difference $>0.24$). In daylight, neighbor rule $N$ flags
$q_{i,t}$ more than 0.32 from the simultaneous city median. Memory rule $M$
uses 24 prior steps (minimum eight, non-daylight residuals zeroed): positive
mean $>0.035$ with $r>0.030$; negative mean $<-0.075$ with $r<-0.060$; or
mean $|r|>0.10$ with current $|r|>0.090$. The composite is
\begin{equation}
 C_{i,t}=B_{i,t}\lor M_{i,t}\lor(T_{i,t}\land N_{i,t})
 \lor(|r_{i,t}|>0.20\land N_{i,t}).
 \label{eq:appendix-composite}
\end{equation}
This batch classifier produces Appendix~B's metrics. The stateful
\texttt{SolarAgent}/\texttt{PlannerAgent} trace instead updates trust,
calibration, participation, and market decisions; the two share physical
evidence but are not the same evaluation object.

\section{Attack-Class Capability and Threshold Sensitivity}
\label{app:attack-sensitivity}
Table~\ref{tab:appendix-attack-capability} reports all eleven scripted classes.
Labels mean strong (S), partial (P), or outside the current claim (O), and do
not guarantee resistance to adaptive attacks.

\begin{table}[H]
  \centering
  \caption{Offline verifier results by attack class. Capability: S=strong, P=partial, O=outside current claim.}
  \label{tab:appendix-attack-capability}
  \normalsize
  \setlength{\tabcolsep}{1.3pt}
  \begin{tabular}{@{}lrrrrl@{}}
    \toprule
    Attack & Rows & P & R & F1 & Cap. \\
    \midrule
    Original FDIA & 1,800 & .976 & 1.000 & .988 & S \\
    Night impossible & 1,440 & 1.000 & 1.000 & 1.000 & S \\
    Above bound & 1,800 & .960 & .984 & .972 & S \\
    Near bound & 1,800 & .000 & .000 & .000 & O \\
    Within-bound fraud & 1,800 & .000 & .000 & .000 & O \\
    Sensor drift & 3,068 & .999 & .864 & .927 & P \\
    Weather spoofing & 1,800 & .000 & .000 & .000 & O \\
    Replay & 2,160 & .992 & .654 & .788 & P \\
    Neighbor incons. & 1,800 & 1.000 & 1.000 & 1.000 & S \\
    Coordinated near bound & 1,945 & .000 & .000 & .000 & O \\
    Intermittent burst & 436 & .990 & .897 & .941 & P \\
    \bottomrule
  \end{tabular}
\end{table}

Strong coverage is obtained for original FDIA, nighttime, above-bound, and
neighbor-inconsistent attacks. Drift, replay, and intermittent bursts are
partial; near-bound, within-bound, weather-spoofing, and coordinated near-bound
attacks remain outside the defense claim.

\begin{table}[H]
  \centering
  \caption{Physics-tolerance sensitivity; $\tau=1.03$ is the implemented point, not a global optimum.}
  \label{tab:appendix-tolerance}
  \normalsize
  \setlength{\tabcolsep}{2.2pt}
  \begin{tabular}{@{}lrrrrrr@{}}
    \toprule
    $\tau$ & Mac. P & Mac. R & Mac. F1 & FRR & Near R & Above R \\
    \midrule
    1.000 & .643 & .574 & .586 & .008 & .572 & .956 \\
    1.005 & .713 & .548 & .592 & .003 & .420 & .955 \\
    1.030 & .545 & .443 & .479 & .000 & .000 & .954 \\
    1.080 & .545 & .426 & .466 & .000 & .000 & .945 \\
    \bottomrule
  \end{tabular}
\end{table}
Raising $\tau$ removes normal false rejections but sacrifices near-bound
recall; above-bound recall is comparatively stable. All points reuse the same
fixed-seed records and scripted attacks.

\section{Allocation Robustness and Audit Evidence}
\label{app:allocation-audit}
The 41-point ratio sweep uses 0.025 increments and seven perturbations: low,
normal, and high demand ($0.5/1.0/1.5$); cloudy generation ($0.6$); 10\% and
20\% FDIA loss ($0.9/0.8$ accepted supply); and high volatility
($1+0.25\sin(0.31t)$ demand).

Within each scenario,
$S=0.50F+0.20R+0.15L+0.10U+0.05D$, with adverse terms reverse-scaled; the
robust score is $0.75\overline{S}+0.25\min(S)$. Feasibility requires average
fulfillment $\geq0.90$, worst fulfillment $\geq0.80$, average slippage
$\leq40\%$, and average producer reward $\geq5{,}000$ kWh.

\begin{table}[H]
  \centering
  \caption{Selected 20/80 result across seven scenarios.}
  \label{tab:appendix-selected-ratio}
  \normalsize
  \setlength{\tabcolsep}{4pt}
  \begin{tabular}{@{}lr@{}}
    \toprule
    Metric & Value \\
    \midrule
    Robust score & .628 \\
    Average/worst scenario score & .651/.556 \\
    Average/worst fulfillment & .956/.836 \\
    Average reward; slippage & 8,976 kWh; 32.569\% \\
    Average unmet demand & 2.111 MWh \\
    \bottomrule
  \end{tabular}
\end{table}
The feasible reward-share interval is 12.5\%--22.5\%; 20\% has the highest
robust score within that set but remains conditional on the simulator,
normalization, scenarios, and weights.

\begin{table}[H]
  \centering
  \caption{End-to-end audit evidence and its boundary.}
  \label{tab:appendix-audit-evidence}
  \normalsize
  \setlength{\tabcolsep}{3pt}
  \begin{tabularx}{\columnwidth}{@{}lXX@{}}
    \toprule
    Layer & Evidence & Boundary \\
    \midrule
    Inspect & Weather, registry & Modeled input, not panel telemetry. \\
    Verify & Report, result & Replays decisions, not sensor truth. \\
    Coordinate & Planner, feedback & Ordered state, not policy optimality. \\
    Allocate & Market, ratio sweep & Declared scenarios and weights. \\
    Settle & Settlement; 5/5 EVM & Consistency, not a security audit. \\
    \bottomrule
  \end{tabularx}
\end{table}

To verify the audit trail, the replay procedure canonicalized each JSON
payload, blanked its \texttt{record\_hash} field, recomputed the SHA-256
digest, and checked the resulting predecessor link. All 144,721 events
across six event types passed this verification. The local EVM suite then
evaluated the allocation workflow under five cases, covering the 20/80
initialization, authorized and unauthorized updates, enforcement of the
10,000-basis-point bound, and settlement after a ratio change. These results
support the internal consistency and replayability of the implementation,
but they do not validate sensor inputs, establish economic optimality, or
constitute a security audit of the contract. Reproduction begins by
identifying the benchmark through its provenance record and generation
digest. The released capability, tolerance, and ratio files can then be
evaluated before replaying the event log and running the EVM tests. Values
reported in the paper are rounded for readability, whereas all verification
procedures use the unrounded records.

\makeatletter
\global\@ACM@balancefalse
\makeatother

\section{User Experience Workflow}
\label{appx:solar-ux}

\proj is designed as a human-in-the-loop energy planning workflow. The user experience begins from a
map-centered dashboard and then moves through verification, registration,
settlement, and audit. Each stage exposes both the user action and the evidence
produced by the system, so that planners can trace how a PV record
becomes an on-chain market state.

\noindent\textbf{Step 1: Wallet and city-state initialization.}
The planner, PV owner, or factory owner first connects a Web3 wallet. After the
signature identity is available, the frontend loads the deployed contract
addresses, ABIs, city layers, PV nodes, factory nodes, and local dashboard state.
This step binds later approvals and transactions to a cryptographic account.

\noindent\textbf{Step 2: Spatial inspection of candidate assets.}
The planner opens the map workspace and inspects distributed energy resources across the demonstration cities. For each candidate PV node, the interface shows
location, panel specifications, battery temperature, DC power, AC power, and predicted generation. This spatial view is important because \proj treats energy generation as a location-dependent urban process rather than as a generic token balance.

\noindent\textbf{Step 3: Physics-bounded verification.}
Before a PV record is allowed to affect the market, the planner reviews the off-chain verification evidence. The physics engine combines geospatial coordinates, irradiance, temperature, panel area, efficiency, and temperature coefficient to compute \(P_{\max}\). If the reported generation satisfies
\(P_{\mathrm{reported}} \leq \tau P_{\max}\), the record is marked \texttt{verified}; otherwise it is labeled \texttt{rejected}. This step blocks FDIA records before they can create false market supply or invalid token
rewards.

\begin{figure}[h]
\centering
\IfFileExists{solar-data-visualization-form.png}{%
\includegraphics[width=\columnwidth]{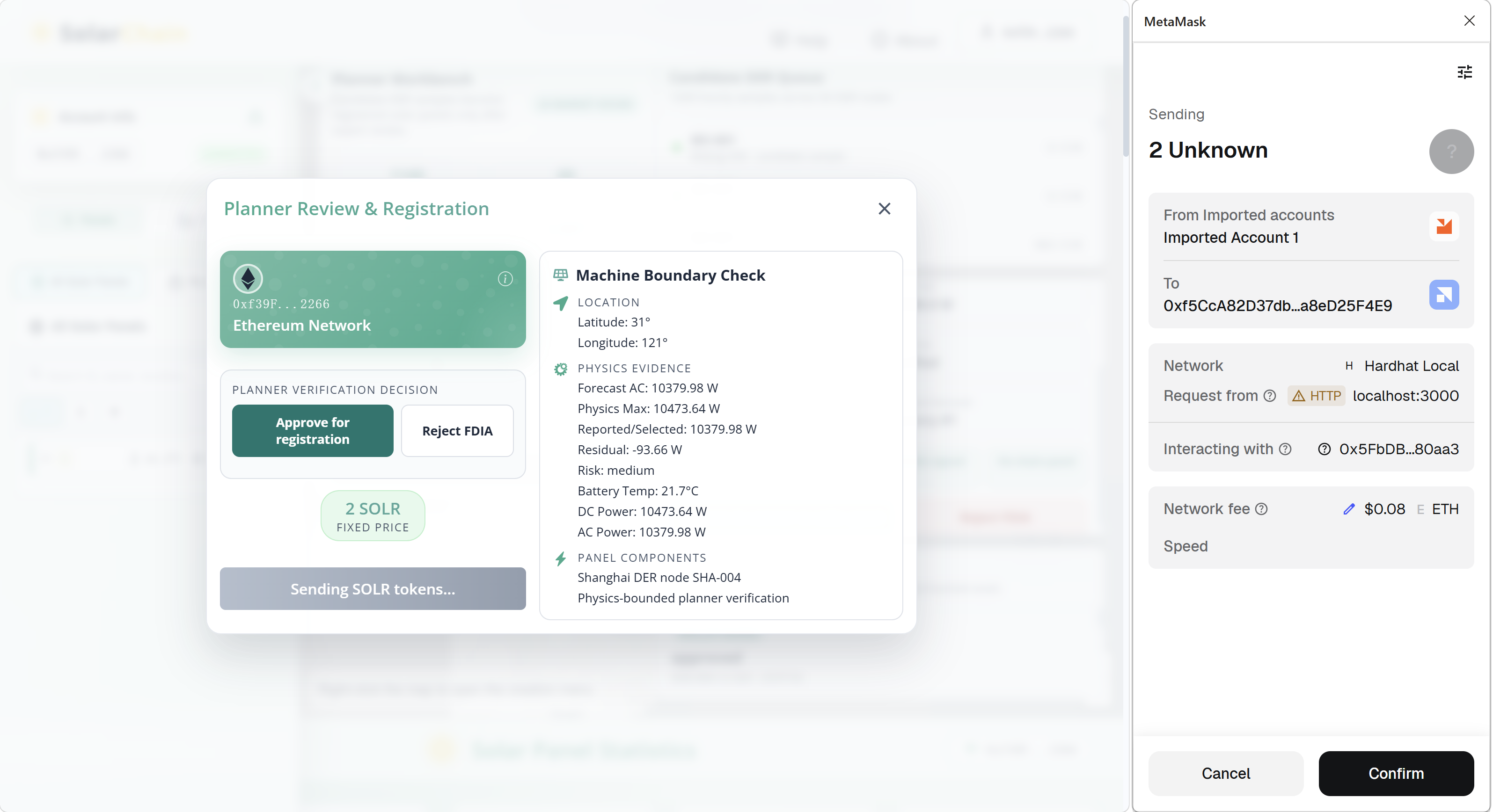}%
}{%
\fbox{\parbox[c][0.22\textheight][c]{0.8\columnwidth}{\centering solar-data-visualization-form.png}}%
}
\caption{Solar panel confirmation. User verifies the selected location and
predicted outputs before submitting a Web3 wallet signature.}
\Description{\proj panel confirmation interface showing selected location and predicted output before wallet signature.}
\label{fig:solar-panel-confirmation}
\end{figure}

\noindent\textbf{Step 4: PV digital-twin registration.}
After a panel passes verification, the PV owner submits the asset to the chain. The panel-registration function records the owner address, location, battery
temperature, DC power, AC power, and creation timestamp. The resulting panel event provides an audit trail from the planner-approved physical record to the
on-chain digital twin. The confirmation view in
Figure~\ref{fig:solar-panel-confirmation} is the final human checkpoint before the wallet signature: the user verifies the selected location and predicted outputs
instead of signing a transaction from an abstract contract call alone.

\noindent\textbf{Step 5: Hourly market-state update.}
At each simulation step, verified generation is aggregated together with factory demand. The market operator calls the market-update function, which applies
\proj's forced distribution rule: 80\% of verified value is added to global market liquidity, while 20\% is assigned to personal producer rewards. This
prevents the marketplace from depending only on voluntary liquidity provision.

\noindent\textbf{Step 6: Reward preview and claiming.}
PV owners can preview their pending reward balance before claiming. If the cooldown interval has passed, the reward contract transfers SOLR to the owner
and updates the last-claim timestamp. This gives owners a visible maintenance incentive while preventing repeated claims within the same simulator interval.

\begin{figure}[h]
\centering
\IfFileExists{energy-exchange-purchase.jpg}{%
\includegraphics[width=\columnwidth]{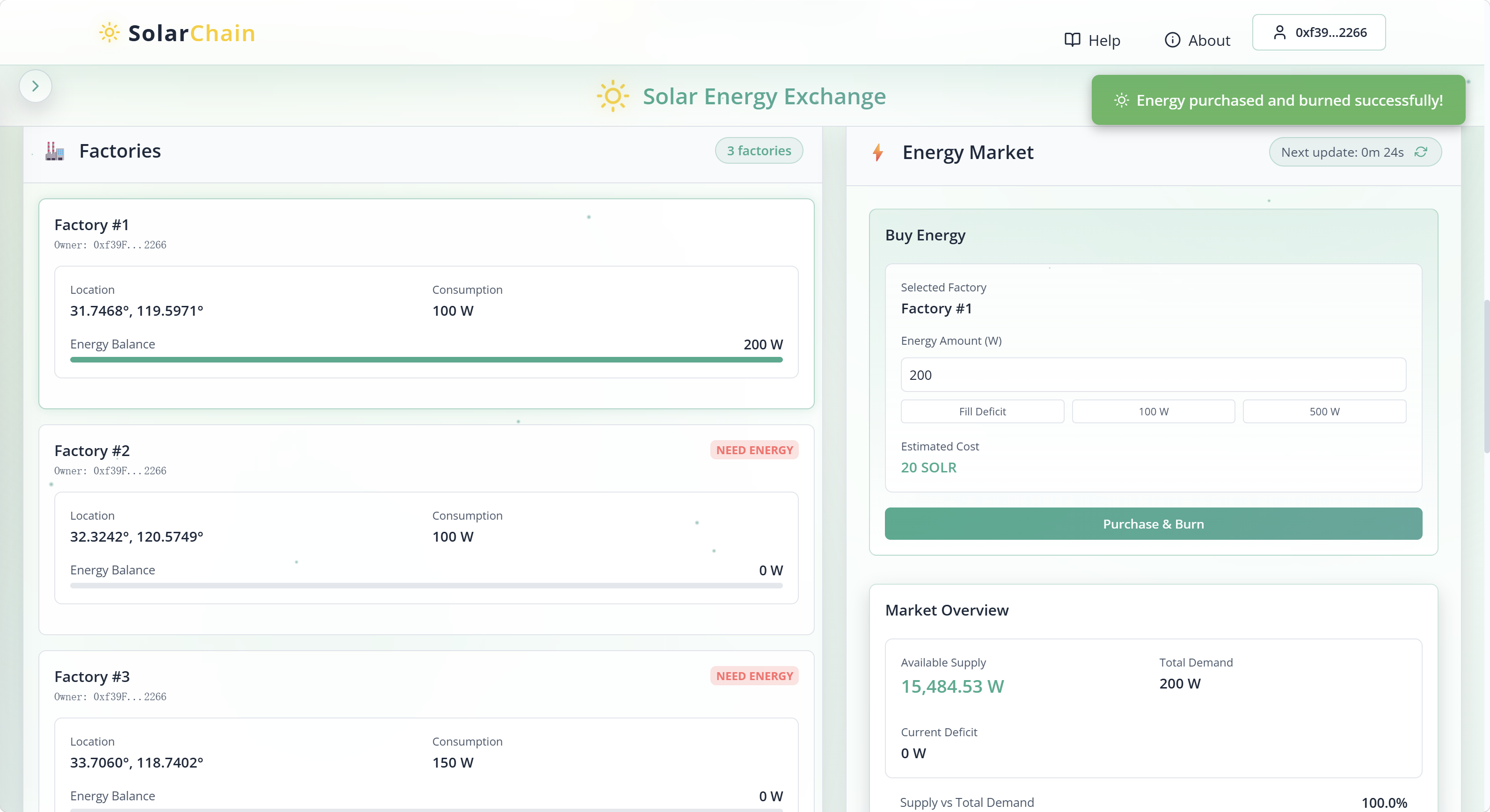}%
}{%
\fbox{\parbox[c][0.22\textheight][c]{0.8\columnwidth}{\centering energy-exchange-purchase.jpg}}%
}
\caption{Energy exchange interface. The user selects a factory, previews the
energy amount, and prepares the SOLR burn transaction.}
\Description{\proj energy exchange interface where a user selects a factory and previews a SOLR burn transaction.}
\label{fig:energy-exchange-purchase}
\end{figure}

\noindent\textbf{Step 7: Factory registration and energy purchase.}
Factory users register demand nodes with location and power-consumption metadata. When buying verified energy, the factory owner approves SOLR spending
and calls the purchase function. The exchange burns the corresponding SOLR, decreases global supply, increases the factory energy balance, and emits an \texttt{EnergyPurchased} event. This ties digital token deflation to energy consumption. Figure~\ref{fig:energy-exchange-purchase} shows the purchase interface where the user selects a factory, previews the energy amount, and prepares the SOLR burn transaction before wallet approval.

\noindent\textbf{Step 8: P2P panel marketplace.}
\proj also supports secondary ownership transfer. A PV owner can list a panel in \texttt{Shop}; buyers place offers after approving token spending; the seller approves one buyer. The shop transfers SOLR from buyer to seller and calls the panel ownership-transfer function, so the asset registry and payment record remain synchronized.

\noindent\textbf{Step 9: Planner audit and analytics.}
Finally, planners review charts for verified generation, rejected anomalies, market liquidity, baseline slippage, token burn, and exergy dissipation. Then dashboard therefore connects meteorological input, physical verification, contract events, and economic settlement into one inspectable evidence chain.

\begin{table*}[t]
\centering
\caption{Detailed data dictionary for released \proj datasets.}
\label{tab:solar-dictionary-short}
\scriptsize
\begin{tabular}{p{0.15\textwidth}p{0.20\textwidth}p{0.59\textwidth}}
\hline
Dataset & Dataset role & Variables: unit; range/domain; frequency; sample observations \\
\hline
\texttt{urban\_energy\_nodes} &
Static PV asset registry. This file defines the physical nodes that later appear
in the generation and verification records. It is used to reproduce geospatial
placement, panel capacity, efficiency assumptions, and temperature response. &
\texttt{node\_id}: id; 50 unique nodes; static node-level; \texttt{BEI-001}.
\texttt{city}: name; Beijing/Shanghai/Chengdu/Shenzhen/Hangzhou; static; Beijing.
\texttt{latitude}: deg.; 22.465945--39.982854; static; 39.887642.
\texttt{longitude}: deg.; 104.022990--121.522048; static; 116.380950.
\texttt{panel\_area\_m2}: m\(^2\); 18.43--63.57; static; 43.39.
\texttt{efficiency}: ratio; 0.1768--0.2234; static; 0.2055.
\texttt{temp\_coefficient}: per \(^{\circ}\)C; -0.00460 to -0.00326; static; -0.00379.
\texttt{install\_date}: date; 2020-01-09--2024-05-10; static; 2022-03-29. \\
\hline
\texttt{spatiotemporal\_generation} &
Hourly physics-verification dataset. This file is the main FDIA benchmark:
each row pairs weather input and panel location with \(P_{\max}\), reported
power, anomaly status, and the final verification decision. &
\texttt{timestamp}: time; 2026-05-01 hourly; node-hour; \texttt{2026-05-01T00:00:00+08:00}.
\texttt{hour}: hour; 0--23; node-hour; 0.
\texttt{node\_id}: id; 50 nodes; node-hour; \texttt{BEI-001}.
\texttt{city}: name; five cities; node-hour; Beijing.
\texttt{latitude}/\texttt{longitude}: deg.; node coordinate ranges above; repeated node-hour; 39.887642/116.380950.
\texttt{irradiance\_Wm2}: W/m\(^2\); 0.00--929.00; node-hour; 15.00.
\texttt{air\_temp\_C}: \(^{\circ}\)C; 9.10--27.00; node-hour; 20.00.
\texttt{P\_max\_W}: W; 0.00--12325.98; node-hour; 138.51.
\texttt{P\_reported\_W}: W; 0.00--13700.91; node-hour; 963.02.
\texttt{fdia\_detected}: bool; True/False, 60 injected FDIA rows; node-hour; False.
\texttt{verification\_status}: label; verified/rejected; node-hour; verified. \\
\hline
\texttt{market\_liquidity} &
Hourly market-state dataset. This file evaluates whether verified generation
improves market depth by comparing \proj liquidity against a no-split
baseline and reporting the corresponding slippage. &
\texttt{timestamp}: time; 24 hourly records; market-hour; \texttt{2026-05-01T00:00:00+08:00}.
\texttt{hour}: hour; 0--23; market-hour; 0.
\texttt{total\_verified\_MW}: MW; 0.000000--0.274561; market-hour; 0.004729.
\texttt{\proj\_liquidity\_MW}: MW; 0.018000--0.270596; market-hour; 0.022350.
\texttt{baseline\_liquidity\_MW}: MW; 0.008000--0.175482; market-hour; 0.010885.
\texttt{slippage\_\proj\_pct}: percent; 0.5703--2.8571; market-hour; 2.6726.
\texttt{slippage\_baseline\_pct}: percent; 1.5235--8.6111; market-hour; 7.9723. \\
\hline
\texttt{p2p\_trades} &
Event-level settlement dataset. This file records simulated factory purchases
during daylight hours and links each trade to energy volume, SOLR burn, and
exergy dissipation for carbon-accounting analysis. &
\texttt{trade\_id}: id; 42 unique trades; event-level; \texttt{TRD-0001}.
\texttt{timestamp}: time; 06:00--19:00 daylight records; event; \texttt{2026-05-01T06:00:00+08:00}.
\texttt{hour}: hour; 6--19; event; 6.
\texttt{factory\_id}: id; six factories; event; \texttt{FAC-BJ-01}.
\texttt{city}: name; five cities; event; Beijing.
\texttt{energy\_purchased\_MW}: MW; 0.000326--0.037502; event; 0.000970.
\texttt{tokens\_burned}: SOLR; 0.3164--38.8174; event; 0.9507.
\texttt{exergy\_dissipated\_MJ}: MJ; 0.0321--4.5783; event; 0.1199. \\
\hline
\end{tabular}
\end{table*}

\section{Dataset Data Dictionary}
\label{appx:solar-data}

The released datasets form a linked physical-to-digital benchmark. The node registry defines each PV asset's location and hardware parameters; the hourly generation file expands those nodes into weather-bounded verification records with FDIA labels. The market-liquidity file aggregates verified generation into liquidity and slippage variables. Additionally, the P2P trade file records factory purchases, SOLR burn, and exergy dissipation. All timestamps use ISO 8601 format in the \texttt{Asia/Shanghai} timezone. To provide a comprehensive overview, Table~\ref{tab:solar-dictionary-short} presents the complete data dictionary, detailing the specific role of each dataset alongside its variable units, numerical ranges or categorical domains, temporal frequencies, and sample observations.

\section{Smart Contracts in \proj}
\label{appx:solar-contracts}

The on-chain layer uses a modular design. \proj keeps the physics model off-chain. An off-chain engine verifies energy generation records, while smart contracts manage state transitions that require public auditing. These transitions include token issuance, asset registration, market allocation, and asset transfers. This design separates expensive physical computations from transparent market settlements.

\begin{table}[htbp]
\centering
\caption{Smart contracts in \proj}
\label{tab:contracts}
\scriptsize
\setlength{\tabcolsep}{3pt}
\renewcommand{\arraystretch}{0.92}
\begin{tabularx}{\columnwidth}{@{}llX@{}}
\toprule
\textbf{Contract Name} & \textbf{Object} & \textbf{Main Function} \\ \midrule
SolarToken      & SOLR token   & Token minting, transfer, approval, and burning \\
SolarPanels     & PV asset     & Panel registration, query, and ownership transfer \\
FactoryRegistry & Demand node  & Factory registration and demand recording \\
EnergyExchange  & Energy market& Supply update, demand tracking, rewards, and purchase \\
PowerReward     & Reward pool  & Power-based reward preview and claiming \\
Shop            & Asset market & Panel listing, offer, sale approval, and transfer \\
\bottomrule
\end{tabularx}
\normalsize
\end{table}

Table~\ref{tab:contracts} summarizes the six core contracts. How do they work together to connect physical data with digital records? As shown in Code box 2, the panel and factory contracts form the registry layer. They do not handle energy trading directly. Instead, they assign stable identities to supply-side solar panels and demand-side factories. This is important because urban energy generation and consumption vary by location. The registry ensures every market event traces back to a specific owner and physical location.

\texttt{SolarToken} and \texttt{EnergyExchange} form the settlement layer. In our system, the SOLR token is an accounting tool, not a speculative asset. Code box 1 outlines its logic, which manages token issuance and verifiable burns. The exchange contract receives verified energy data from the off-chain engine. Code box 3 demonstrates how the system divides this energy into two parts: market liquidity and owner rewards. When a factory buys energy, it does not simply pay another user. Instead, the contract burns the spent SOLR tokens. This reduces the global token supply and increases the factory's energy balance. This mechanism directly connects digital token deflation to physical energy consumption.

Finally, the reward and shop contracts encourage long-term participation. The reward contract distributes tokens based on panel capacity. It uses a cooldown timer to prevent multiple claims in the same simulation step. The shop contract manages the secondary market for solar panels. It verifies ownership, handles payments, and updates the registry (Code box 2). Together, these contracts tie economic incentives to physical infrastructure rather than short-term token trading.

\begingroup
\setlength{\fboxsep}{2pt}
\setlength{\parskip}{0pt}
\vspace{4mm}
\noindent\textbf{Code box 1: token issuance and burn.}\par
\noindent\fbox{\begin{minipage}{0.97\columnwidth}
\tiny\ttfamily
mint(to, amount): require supply + amount <= cap;\\
\hspace*{1em}\_mint(to, amount).\\
burn(amount): \_burn(msg.sender, amount).\\
burnFrom(account, amount): spend allowance;\\
\hspace*{1em}\_burn(account, amount).
\end{minipage}}\par\vspace{0.5ex}
\vspace{4mm}
\noindent\textbf{Code box 2: registry and ownership evidence.}\par
\noindent\fbox{\begin{minipage}{0.97\columnwidth}
\tiny\ttfamily
createPanel(lat, lon, temp, dcPower, acPower):\\
\hspace*{1em}panelCount++; save owner, location, power.\\
createFactory(lat, lon, powerConsumption):\\
\hspace*{1em}factoryCount++; save owner, location, demand.\\
approveSale(itemId, buyer): transfer SOLR;\\
\hspace*{1em}transferPanelOwnership(panelId, buyer).
\end{minipage}}\par\vspace{0.5ex}
\vspace{4mm}
\noindent\textbf{Code box 3: forced liquidity, rewards, and consumption burn.}\par
\noindent\fbox{\begin{minipage}{0.97\columnwidth}
\tiny\ttfamily
updateMarketStep(users, userEnergy, totalEnergy, demand):\\
\hspace*{1em}reward = userEnergy[i] * 20 / 100;\\
\hspace*{1em}personalRewardWei[user] += reward * 1e18 / 100000;\\
\hspace*{1em}globalSupplyEnergy += totalEnergy * 80 / 100;\\
\hspace*{1em}totalDemandEnergy = demand.\\
buyEnergyForFactory(factoryId, energyAmount):\\
\hspace*{1em}require supply >= energyAmount;\\
\hspace*{1em}costWei = energyAmount * 1e18 / 100000;\\
\hspace*{1em}burnFrom(msg.sender, costWei);\\
\hspace*{1em}supply -= energyAmount; factory balance += energy.
\end{minipage}}\par

\endgroup

\end{document}